\documentclass[prd,aps,showpacs,nofootinbib,preprint,eqsecnum]{revtex4}

\usepackage{amsmath}
\usepackage{amssymb}
\usepackage{amsfonts}
\usepackage{graphicx,bm}
\usepackage{dcolumn}

\usepackage{color,amsxtra}
\usepackage{epsf}

\usepackage{enumerate}
\usepackage{hhline}
\usepackage{array}
\usepackage{tabularx}

\usepackage{subfigure}

\usepackage{fancyhdr}
\usepackage{mathrsfs}

\usepackage{amsmath}
\usepackage{amsthm}
\usepackage{amssymb}
\usepackage{amscd}
\usepackage[british]{babel}
\usepackage{babel}
\usepackage{graphicx}
\usepackage{psfrag}
\usepackage{epsfig}
\usepackage{rotating}
\usepackage{times}

\theoremstyle{plain}

\newcommand{\boxend}{\flushright{$\Box$}}

\begin{document}

\title{On the viability of quintessential inflation models from observational data }

\author{ Jaume de Haro$^{1}$\footnote{E-mail: jaime.haro@upc.edu}}

\affiliation{$^{1}$Departament de Matem\`atica Aplicada, Universitat
Polit\`ecnica de Catalunya, Diagonal 647, 08028 Barcelona, Spain}

\thispagestyle{empty}

\begin{abstract}
{Assuming that primordial density fluctuations are nearly Gaussian, from a frequentist viewpoint,
the two-dimensional marginalized joint coincidence contour in the plane $(n_s,r)$ (being $n_s$ the spectral index and $r$ the ratio of tensor to scalar perturbations), without
the presence of running is often used to test the viability of the inflationary models.
The models that provide, between $50$ and $60$ e-folds, a curve in that plane lying outside the $95.5 \%$ C.L are ruled out. I
will basically argue that, in quintessential inflation,  this low number of e-folds is unjustified, and that  models leading to a theoretical value of the running different from zero must  be checked
with observational data allowing the running. When both prescriptions are taken into account, dealing in the context of
quintessential inflation, i.e. when the potential is a combination of an inflationary with a quintessential one that leads to a {kination (also called deflation)} regime,
inflationary models such as the quartic
or the Higgs potential
are allowed.
}
\end{abstract}


\maketitle

\section{ Introduction}

The inflationary paradigm is an implementation of General Relativity  introduced by A. Guth in his seminal paper \cite{guth} to solve the well-known flatness and horizon problems and refined by A. Linde,  A. Albrecht and P.J. Steinhardt  \cite{linde,lindea,steinhardt}. Later, some authors  such as V. Mukhanov , A. Starobinsky  or  A. Guth himself \cite{Mukhanov, Mukhanova, starobinsky,gutha,bardeen,sasaki} realize that inflation could explain, via quantum fluctuations, the origin of primordial inhomogeneities. In the first and simplest version, inflation was driven by a single scalar field, named inflation,  minimally coupled with gravity
and where a potential had a deep well in order that the inflation field could oscillate and thus, release its energy creating particles that reheat the Universe and matching  with the standard hot Friedmann Universe.  However,  soon after several authors  introduced multi-fields to implement inflation \cite{starobinskya,turnera,la, barrow, steinhardta, bellido, lindeb, adams, copeland} , and calculate, in that case, the relevant quantities  coming from perturbations theory \cite{sasakia,wandsa}. At the same time, theories beyond General Relativity such as $f(R)$ gravity , tensor-scalar theories 
 or the combination of both  were used to introduce this inflationary period  \cite{starobinskyb,bellidoa,wands,wandsb}.  
 
 The paradigm changed when at the end of last century
 observations of distant type-Ia supernovae, baryonic acoustic oscillations, anisotropies of the cosmic microwave background radiation, and some other data confirmed the current cosmic acceleration \cite{perlmutter,perlmuttera,riess}, then some new models were developed to understand this behavior and unify the early acceleration with the late acceleration, such as the introduction of the cosmological constant \cite{armendariz}, quintessence models \cite{steinhardtb,steinhardtc},  new models of $f(R)$ gravity \cite{odintsova,odintsovb,odintsovc}, and more recently the combination of $f(R)$ gravity with tensor-scalar theories \cite{odintsovd}.
However, in spite of inflation being the most popular solution, there are other scenarios such as  the bouncing cosmologies, where the initial singularity was replaced by a bounce,
that overpass  the problems associate to  the standard Big Bang theory \cite{novello, brandenberger}. In fact, the most promising alternative to the inflationary paradigm is the so-called matter bounce scenario in Loop Quantum Cosmology  and its refinements \cite{wilson,amoros, wilsona,amorosa}, where in the contracting phase the Universe is matter dominated and after the bounce it matches with the standard Friedmann model.

Nowadays,
recent  Planck's observational data are used to check the viability of theoretical inflationary models (see for instance \cite{odintsove, odintsovf,kaiser}). In the frequentist analysis, usually two essential prescriptions are used
to do that task.
The first one, is related with the number of e-folds from observable scales exiting the Hubble radius towards the end of inflation. It is generally assumed that this number
ranges from $50$ to $60$ e-folds \cite{Ade}, but this assumption comes from the fact that previously it was imposed that from the end of inflation to the beginning of  reheating,
the universe was matter dominated \cite{ll}. However, it is well {}{known} that different potentials lead to different phases during this period, and as we will see, an Equation of
State (EoS) parameter greater than zero during this phase increases the number of e-folds. Therefore, to be rigorous with the analysis, the number
of e-folds must be model dependent.

The other prescription is related with the two dimensional marginalized analysis en the plane $(n_s,r)$, where $n_s$ is the spectral index and $r$ is the ratio of tensor
to scalar perturbations. Since each models leads to a curve in this plane that could be parametrized by the number of e-folds, when the piece of it that contains the number
of allowed e-folds belongs in the $95.5 \%$ C.L., the model is viable. However, there is a great difference whether the presence of running is allowed or not, because the
running increases the area of the region that contains the $2\sigma$ C.L.. Thus, when the running is allowed, it is easier for a given inflationary  model to pass the test than when the
running is not taken into account. However, although the theoretical inflationary model leads, in general, to a running of the spectral index, their curves in
the plane $(n_s,r)$ are usually compared with the observational $95.5 \%$ C.L. without the presence of running, which rules out  some models that ought not to be disregarded,
such as
 {}{quintessential}   models where inflation is given by a quartic  or a one-dimensional Higgs potential.

The paper is organized as follows: In section 2 I review the calculation of e-folds as a function of the reheating temperature and the EoS parameter of the
effective fluid that drives the universe from the end of inflation to the beginning of the radiation era. In section 3,
I propose some  {}{quintessential inflation} models whose
potential is the combination of the {}{Higgs-style} potential and the cosmological constant -note that, recently the Higgs field was proposed as responsible, solely
or in part, for inflation \cite{trashorras,figueroa}-, {}{and others that are the
combination of a quartic one and  a cosmological
 constant}.
{}{Section 4} is devoted to the study of the reheating in such as models. We will see that due to the phase transition the gravitational production of
heavy massive particles with a mass about {}{$10^{13}$ GeV,} could reheat the universe at temperatures of the order of $10$ GeV. In last section, I perform
a detailed calculation of the number of e-folds for that {}{non oscillating} models showing its viability form the fact that the
number of e-folds is always greater than $65$.

The units used throughout the paper are $\hbar=c=1$.

\section{The number of e-folds}

To calculate the number of e-folds from the exiting of a pivot scale $k_*$ to the end of the inflation, I use the well-known formula \cite{ll}
\begin{eqnarray}
\frac{k_*}{a_0H_0}
=e^{-N(k_*)}\frac{H(k_*)}{H_0}\frac{a_{end}}{a_R}\frac{a_R}{a_M}\frac{a_M}{a_0},
\end{eqnarray}
where $R$ (resp. $M$) denotes the point when radiation (resp. matter) starts to dominate,
the sub-index $0$ refers to the present time and $end$ denotes the end of the inflation.

Now we have to relate the scale factor with the corresponding energy density. From the end of inflation to the beginning of the reheating I will assume that  the
 EoS parameter, namely $w$, is constant. Then, one has
\begin{eqnarray}
\left(\frac{a_{end}}{a_R}\right)^{3(1+w)}=\frac{\rho_R}{\rho_{end}}, \quad
\left(\frac{a_R}{a_M}\right)^4=\frac{\rho_M}{\rho_R},
\end{eqnarray}
 and consequently, if we take as a pivot scale the same as in \cite{Ade}, i.e.,  $k_*=0.05$ $\mbox{Mpc}^{-1}$, since the current horizon scale is
 $a_0H_0\cong 2\times 10^{-4}$ $\mbox{Mpc}^{-1}$ (where, as usual, I choose $a_0=1$) , one will obtain
\begin{eqnarray}
N(k_*)=-5.52+\ln\left(\frac{H(k_*)}{H_0} \right)+
\frac{1}{4}\ln\left(\frac{\rho_M}{\rho_R} \right)
+\frac{1}{3(1+w)}\ln\left(\frac{\rho_R}{\rho_{end}} \right)
+\ln\left(\frac{a_M}{a_0} \right).
\end{eqnarray}

Writing  this expression as follows
\begin{eqnarray}\label{N}
N(k_*)=\left\{-5.52+\ln\left(\frac{H(k_*)}{H_0} \right)+
\frac{1}{4}\ln\left(\frac{\rho_M}{\rho(k_*)} \right)
+\ln\left(\frac{a_M}{a_0} \right)\right\}\nonumber\\
+\left\{\frac{1}{4}\ln\left(\frac{\rho(k_*)}{\rho_R} \right)
+
\frac{1}{3(1+w)}\ln\left(\frac{\rho_R}{\rho_{end}} \right)\right\},
\end{eqnarray}
and
using that the temperature of the universe at the beginning of the matter domination era, namely $T_M$, could be calculated from
the formula $\rho_M\cong \frac{\pi^2}{15}g_M T_M^4$ with $g_M\cong 3.36$ \cite{rg}, and that the process
is adiabatic after reheating, i.e., $T_0=\frac{a_M}{a_0}T_M$, we can write {}{
\begin{eqnarray}
N(k_*)\cong\left\{-5.52+\frac{1}{2}\ln\left(\frac{H(k_*)}{M_{pl}} \right)+\frac{1}{4}\ln\left(\frac{\pi^2}{45}g_M  \right)
+
\ln\left(\frac{T_0}{H_0} \right)\right\}\nonumber\\
+\left\{\frac{1}{4}\ln\left(\frac{\rho(k_*)}{\rho_{end}} \right)
+\left[\frac{1}{3(1+w)}
-\frac{1}{4}\right]\ln\left(\frac{\rho_R}{\rho_{end}} \right)\right\}
.
\end{eqnarray}
}

To evaluate this quantity we need the following observation data: $T_0\cong 2.73 $ K $\cong 2.34\times 10^{-13}$ GeV, $H_0\sim 6\times 10^{-61} M_{pl} \cong 1.46\times 10^{-42}$ GeV, and
\begin{eqnarray}\label{power1}
 {\mathcal P}\cong \frac{H^2(k_*)}{8\pi^2M_{pl}^2\epsilon}
 \cong 2\times 10^{-9}
\end{eqnarray}
where $\epsilon\cong\frac{M_{pl}^2}{2}\left(\frac{V_{\varphi}}{V} \right)^2$ is the main slow roll parameter (see for instance \cite{bld}).

If ones assumes, as in \cite{ll}, that there is not substantial drop of energy at last stages of inflation ($\rho(k_*)\cong \rho_{end}$) one obtains
\begin{eqnarray}\label{C}
N(k_*)\cong 58+\frac{1}{4}\ln\epsilon+\left[\frac{1}{3(1+w)}-\frac{1}{4}\right]\ln\left(\frac{\rho_R}{\rho_{end}} \right).
\end{eqnarray}

From this last formula we see the importance of the EoS parameter between the end of the inflation and the beginning of the reheating. For example, for power law potentials
$V(\varphi)= V_0\left(\frac{\varphi}{M_ {pl}}\right)^{2n}$, where reheating is due to the oscillations of the inflaton, one has $w\cong \frac{n-1}{n+1}$ \cite{turner,ford}, and thus,
using $\rho_R\cong \frac{\pi^2}{30}g_R T_R^4 $ one gets
\begin{eqnarray}
N(k_*)\cong 58+\frac{1}{4}\ln\epsilon+\frac{2-n}{12n}\ln\left(\frac{\rho_R}{\rho_{end}} \right)
\cong
58+\frac{1}{4}\ln\epsilon
\nonumber\\
+\frac{2-n}{3n}\left[\frac{1}{4}\ln\left(\frac{\pi^2}{30}g_R \right)  +\ln\left(\frac{T_R}{\rho_{end}^{\frac{1}{4}}} \right)\right],
\end{eqnarray}
{}{which} means that the last term is negative, and the number of e-folds  decreases with the {}{reheating temperature},  only when $n<2$.

Now taking into account that for power law potentials 
one has (see page $10$ of \cite{Ade}, where one has to replace $n$ by $2n$, because  in \cite{Ade} the authors consider the  power law potential 
$V(\varphi)= V_0\left(\frac{\varphi}{M_ {pl}}\right)^{n}$ instead of the potential $V(\varphi)= V_0\left(\frac{\varphi}{M_ {pl}}\right)^{2n}$ considered here) 
\begin{eqnarray}
n_s-1=-4n(n+1)\frac{M_{pl}^2}{\varphi^2},\qquad r=16\epsilon=32 n^2\frac{M_{pl}^2}{\varphi^2},\end{eqnarray}
one can deduce that 
the slow roll parameter $\epsilon$ is related with the spectral index via the formula $\epsilon=\frac{n(1-n_s)}{2(n+1)}$, and
since I am assuming
{}{$\rho_{end}\cong \rho(k_*)\cong 48\pi^2\epsilon \times 10^{-9} M_{pl}$,} we will obtain the formula
\begin{eqnarray}\label{number}
N(k_*)
\cong
58
+\frac{2n-1}{6n}\ln\left( \frac{n(1-n_s)}{2(n+1)}\right)
+\frac{2-n}{3n}\left[\frac{1}{4}\ln\left(\frac{g_R}{144}\right) +\ln\left(\frac{100 T_R}{M_{pl}} \right)\right].
\end{eqnarray}

We can see that second term in the rhs depends on the spectral index, but  a small perturbation of its value do not  practically change the value of the term  $\frac{2n-1}{6n}\ln\left( \frac{n(1-n_s)}{2(n+1)}\right)$  and consequently the value of the 
number of e-folds is unchanged. Therefore,  I can safely  take its central value $n_s=0.9603$ \cite{Ade}.
As a consequence, for reheating temperatures {}{consistent with the bounds {}{coming} from nucleosynthesis}, i.e., in the
range of $10^9$ GeV and $1$ MeV {}{(see for instance  \cite{gkr,hannestad}, where it is argued that a reheating temperature could be in the MeV regime.
 Moreover,
 this low temperature prevents a late time entropy production due to the decay of non-relativistic gravitational relics such as gravitinos or moduli particles \cite{kks})}
and using that
$g_R
= 107$ for
$T_R
\geq
175$ GeV, $g_R=90$ for
$200$ MeV$\leq T_R
\leq
175$ GeV, and  $g_R=11$ for
$1$ MeV$\leq T_R
\leq
200$ MeV \cite{rg},
one obtains the bound
$42\leq N(k_*)\leq 52$ for a quadratic potential ($n=1$). For a quartic potential ($n=2$), which includes the one-dimensional Higgs potential, one has $N(k_*)\cong 57$,
and finally, for $n>2$ the number of e-folds {}{lies} between
$
 58
+\frac{2n-1}{6n}\ln\left( \frac{n(1-n_s)}{2(n+1)}\right)-17\times\frac{2-n}{3n}$
and 
$58
+\frac{2n-1}{6n}\ln\left( \frac{n(1-n_s)}{2(n+1)}\right)-45\times \frac{2-n}{3n}.
$
The maximum number of e-folds, is obtained taking the limit $n\rightarrow \infty$, which gives
$62\leq N(k_*)\leq 72$.

Note that my result is in complete agreement with the results of \cite{rg}, and differs about $5$ e-folds of the results presented in \cite{ll}, because in that work the pivot
scale is taken for modes at the current
Hubble scales $k_*=a_0H_0$. In fact, the pivot scale appears in equation (\ref{N}) with the term $-\ln\left( \frac{k_*}{a_oH_0}\right)$, which value is $-5.52$ for the pivot scale used by Planck2013 team \cite{Ade}, i.e., for
$k_*=0.05$ $\mbox{Mpc}^{-1}$, and $0$ when $k_*$ coincides with the current horizon scale, i.e., for $k_*=a_0H_0$ which is the value used in \cite{ll}.
If one chooses  $k_*=0.02$ $\mbox{Mpc}^{-1}$, which is the value used by Planck2015 team \cite{P2015},   one has 
$-\ln\left( \frac{k_*}{a_oH_0}\right)=-4.6$. Thus, we can conclude that depending of the pivot scale chosen the final result could change around $5$ e-folds.

On the other hand,
for {}{quintessential inflation}, that is, for non oscillatory  models, one also obtains a number of e-folds greater than $62$ because at the end of inflation the energy density of the background must decay
faster than that of radiation, this means that the number $\frac{1}{3(1+w)}-\frac{1}{4}$ that appear in equation (\ref{C}) must be negative ($1\geq w>\frac{1}{3}$). The maximum of e-folds is obtained when $w=1$
that corresponds to the {}{kination \cite{joyce} (also called deflation \cite{Spokoiny})} regime, where all the potential energy of the field is transformed in  kinetic.

This result is very important in order to disregard inflationary models, because when one considers the two-dimensional marginalized joint coincidence contours in the plane $(n_s, r)$,
begin $n_s$ the spectral index and $r$ the tensor/scalar ratio, one usually  restricts the value of the number of e-folds between $50$ and $60$, this clearly, in the case of polynomial
potentials, is only justified for the quadratic one. Moreover, usually this analysis is done without the presence of running. However the majority of the models discussed {}{have} running, 
then
a more accurate analysis, must allow the running, which extends the area within  confidence limits, consequently establishing the viability of some models.

\begin{figure}[h]
\begin{center}
\includegraphics[scale=0.50]{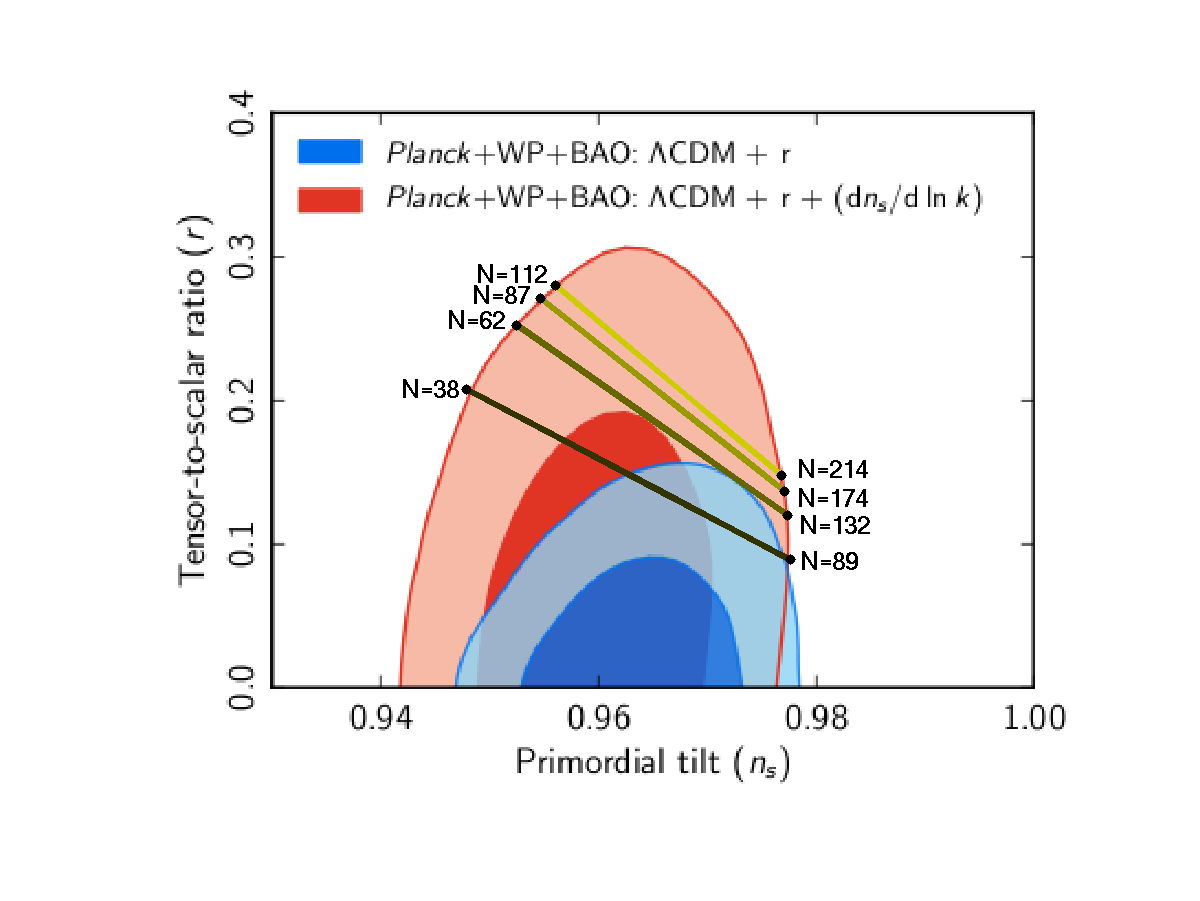}
\end{center}

\caption{ {\protect\small Marginalized joint confidence contours for $(n_{\mathrm s} \,, r)$,
at the 68\,\% and 95\,\% CL, with and without the presence of running of the spectral indices. From darker to lighter, we have plotted the curves $(n_s(N), r(N))$ for
power law potentials with $n=1,2,3$ and $4$.
({Figure courtesy of the Planck2013 Collaboration}).  }}
\end{figure}

Dealing  with power law potentials $V(\varphi)= V_0\left(\frac{\varphi}{M_ {pl}}\right)^{2n}$,
using the well-known formulas  for the slow-roll parameters and the number of e-folds 
 \cite{btw}:
\begin{eqnarray}
\epsilon\cong \frac{M^2_{pl}}{2}\left(\frac{V_{\varphi}}{V} \right)^2,
\quad \eta\cong {M^2_{pl}}\left(\frac{V_{\varphi\varphi}}{V} \right)\quad N\cong \frac{1}{M_{pl}^2}\left|\int_{\varphi}^{\varphi_{end}} \frac{V}{V_{\varphi}}d\varphi\right|,
\end{eqnarray}
 and taking into account the relations
\begin{eqnarray}\label{relations}
n_s-1=-6\epsilon+2\eta,\qquad r=2\epsilon,
\end{eqnarray}
a simple calculation leads to {}{
\begin{eqnarray}\label{A1}
n_s(N)=1-\frac{2n+2}{n+2N},\quad  r(N)=\frac{16n}{n+2N}, 
  \Longrightarrow n_s=1-\frac{n+1}{8n}r.
\end{eqnarray}

We can see in the figure $1$ that for the quadratic potential, in the presence of the running and dealing in the $95.5$ \% C.L. the number of e-folds allowed is from $38$ to $89$,
and thus, in this case the quadratic potential is viable.
However, for the quartic potential the minimum number of e-fold is $62$, which is forbidden because reheating considerations leads to about $57$ e-folds.
The same happens with the one-dimensional Higgs potential \cite{DWI, Vilenkin}, which leads approximately to the same number of e-folds and the curve in the plane $(n_s,r)$ as the
quartic potential. An unusual way to allow the viability of that models is to consider the quartic or Higgs field coupled to gravity with  a very large coupling constant \cite{Higgs}.  Fortunately, as we will see some  {}{quintessential inflation} potentials with an phase transition to a {}{ kination regime \cite{joyce,Spokoiny}},
whose inflationary part is given by a quartic  \cite{pv} or a Higgs potential \cite{hap}, will be allowed.

Finally, for
values of $n$ greater {}{or equal to} $3$, the minimum number of e-folds needed to enter in the $95.5$ \% C.L. is greater than $87$ which is also forbidden by the reheating considerations
presented above.

\section{Quintessential inflation models}
In this section we will present some quintessential {}{inflation} models --a combination of the Higgs potential with a constant--
{}{which are the generalization of the potential obtained in \cite{hap} and
 where the reheating temperature and the number of e-folds could be calculated more accurately,}
because we have an analytic expression of the background.
We start with the following dynamical equation, {}{which depend on two parameters $H_e$ and $b$}
{}{\begin{eqnarray}\label{background}
 \dot{H}=\left\{\begin{array}{ccc}
                 -3H_e(bH-(b-1)H_e)& \mbox{for} & H>H_E\\
                 -3H^2+{\Lambda} & \mbox{for} & H\leq H_E,
                \end{array}\right.
\end{eqnarray}
where
$b\geq 2$, {}{$\Lambda\ll H_e^2$} is a cosmological constant,
$H_E=\frac{bH_e}{2}\left(1-\sqrt{1-\frac{4(b-1)}{b^2}+\frac{4\Lambda}{3b^2H_e^2}}  \right)$,
is the value of the Hubble parameter at the transition time, which
ensures that  its derivative is continuous at the phase transition, and since the cosmological constant is very small, we can see that
$H_e\cong H_E$ is approximately the value of the Hubble parameter at the transition time.}

{}{To obtain the potential which leads to this dynamics, i.e.,
to obtain $V(\varphi)$ whose conservation equation $\ddot{\varphi}+3H\dot{\varphi}+V_{\varphi}=0$ has a solution, namely $\varphi(t)$,
which leads to a Hubble parameter $H(t)=\sqrt{\frac{\dot{\varphi}^2(t)+2V(\varphi(t))}{6M_{pl}^2}}$ satisfying (\ref{background}),
we will use the reconstruction method (see \cite{hap} for details), based in 
the formula
\begin{eqnarray}
 \varphi=M_{pl}\int \sqrt{-2\dot{H}}dt=-M_{pl}\int \sqrt{\frac{-2}{\dot{H}}}dH.
\end{eqnarray}

In the case of our particular model we will obtain for {}{$H>H_E$}
\begin{eqnarray}
\varphi=-2M_{pl}\sqrt{\frac{2H}{3bH_e}-\frac{2(b-1)}{3b^2}}\Longleftrightarrow  H=\left(\frac{3b\varphi^2}{8M_{pl}^2}+\frac{b-1}{b}\right)H_e.
\end{eqnarray}

Finally, using that $V=3H^2M_{pl}^2+\dot{H}M_{pl}^2$, for {}{$\varphi<\varphi_E\equiv \varphi(H_E)$}  one obtains
\begin{eqnarray}
V(\varphi)={\frac{27H_e^2b^2}{64M_{pl}^2}} \left(\varphi^2-\frac{8}{3}\left(\frac{1}{2}-\frac{b-1}{b^2}\right)
M_{pl}^2\right)^2
 -\frac{16(b-2)^2}{9b^2}{\frac{27H_e^2b^2}{64M_{pl}^2}}  M_{pl}^4.
\end{eqnarray}

On the other hand, for {}{$\varphi\geq \varphi_E$} since the universe is driven by an stiff fluid plus a cosmological constant one will obtain
{}{$V(\varphi)=\Lambda M_{pl}^2$}.

Summing up, the potential which leads to the background (\ref{background}) is
\begin{eqnarray}\label{Pot}
V(\varphi)=\left\{\begin{array}{ccc}
\lambda\left(\varphi^2-\frac{8}{3}\left(\frac{1}{2}-\frac{b-1}{b^2}\right)
 M_{pl}^2\right)^2-\gamma& \mbox{for} & \varphi\leq \varphi_E\\
 \Lambda M_{pl}^2 & \mbox{for} & \varphi\geq \varphi_E,
 \end{array}\right.
\end{eqnarray}
where, in order to simplify, we have introduced the notation
$\lambda ={\frac{27H_e^2b^2}{64M_{pl}^2}} $,
$\gamma= \frac{16(b-2)^2}{9b^2}\lambda M_{pl}^4 $ and
$\varphi_E=-2M_{pl}\sqrt{\frac{2H_E}{3bH_e}-\frac{2(b-1)}{3b^2}}\cong-\frac{2}{b}\sqrt{\frac{2}{3}}M_{pl}$ is the value of the field at the transition time.

\vspace{0.5cm}

Note that the the inflationary piece of (\ref{Pot}) is a Higgs-style potential minus the constant   $\gamma= \frac{16(b-2)^2}{9b^2}\lambda M_{pl}^4 $,
which vanishes 
 for $b=2$  obtaining the following exact
Higgs-style potential
\begin{eqnarray}
V(\varphi)=\left\{\begin{array}{ccc}
\lambda\left(\varphi^2-\frac{2}{3}
 M_{pl}^2\right)^2& \mbox{for} & \varphi\leq \varphi_E\\
 \Lambda M_{pl}^2 & \mbox{for} & \varphi\geq \varphi_E.
 \end{array}\right.
\end{eqnarray}

On the other hand, for $b>2$ becomes negative in the interval between
$-\sqrt{\frac{8}{3}\left(\frac{1}{2}-\frac{b-1}{b^2}\right)
 M_{pl}^2+\sqrt{\gamma/\lambda}}$ and $-\sqrt{\frac{8}{3}\left(\frac{1}{2}-\frac{b-1}{b^2}\right)
 M_{pl}^2-
 \sqrt{\gamma/\lambda}}\leq \varphi_E
$, meaning that
immediately before the phase transition the universe enters in a short ekpyrotic regime. After the phase transition the universe is driven by an stiff fluid plus a cosmological
constant (kination phase). However, since in quintessential inflation,  it is usual to consider positive potentials one can choose
a similar potential, for example, a quartic one  with the shape
\begin{eqnarray}\label{Pot1}V(\varphi)=\left\{\begin{array}{ccc}
\bar{\lambda}(\varphi^4-M_{pl}^4)+\Lambda M_{pl}^2 & \mbox{for} & \varphi\leq -M_{pl}\\
\Lambda M_{pl}^2 & \mbox{for} & \varphi\geq -M_{pl},
 \end{array}\right.
\end{eqnarray}
which is always positive.

The difference between  potentials  (\ref{Pot1}) and (\ref{Pot}), is that the the Higgs-style one given by  equation (\ref{Pot}) has, by construction, an analytic
 solution: the background (\ref{background}) that allows us to perform  calculations
such as the reheating temperature or the number of e-folds analytically, and thus,  without any dubiously justified assumption.}

\subsection{Slow roll parameters for the models}

For this model is easy to calculate the slow roll parameters, using the exact formulas \cite{btw}
\begin{eqnarray}
 \epsilon =-\frac{\dot{H}}{H^2}, \quad \eta=2\epsilon-\frac{\dot{\epsilon}}{2H\epsilon}.
\end{eqnarray}

For our model, taking $x\equiv \frac{3H_eb}{H} $ we have
\begin{eqnarray}
 \epsilon=x\left(1-\frac{b-1}{3b^2}x\right), \quad \eta=\epsilon+\frac{x}{2},
\end{eqnarray}
and thus, from the relations (\ref{relations}) we get
\begin{eqnarray}\label{ns vs r}
 n_s=1-3x+\frac{4(b-1)}{3b^2}x^2, \quad r=16x-\frac{16(b-1)}{3b^2}x^2.
\end{eqnarray}

{}{Note that, if we assume, as usual, that inflation ends when the universe starts to decelerate ($\epsilon=1$, i.e., $\dot{H}=-H^2$), we will obtain that, for our model, inflation ends at
\begin{eqnarray}\label{end}
H_{end}=\frac{3H_eb}{2}\left(1+\sqrt{1-\frac{4(b-1)}{3b^2}}   \right)> 3H_e\cong 3H_E.\end{eqnarray}}

Moreover,
to evaluate the value of the parameter $\lambda$ one has to
insert the relation {}{$H^2(k_*)\cong \frac{9H_e^2b^2}{\epsilon^2}\cong \frac{64\lambda}{3\epsilon^2}M_{pl}$ }(recall that when the pivot scale leave the Hubble radius $x\cong \epsilon$)
in formula (\ref{power1}).
Then, since {}{for the potential (\ref{Pot})}
 ones has $\epsilon\cong \frac{1-n_s}{3}$, taking,  $n_s\cong 0.96$, {}{one gets
\begin{eqnarray}\label{lambda}
 \lambda\sim  10^{-14} \Longleftrightarrow H_e\sim \frac{1}{b}10^{-7} M_{pl}.
\end{eqnarray}


On the other hand, for the quartic potential (\ref{Pot1}),
using that $\epsilon\cong\frac{8\varphi^6M_{pl}^2}{(\varphi^4-M_{pl}^4)^2}$
a simple calculation shows that inflation ends when $\varphi_{end}\cong \sqrt{8.25}M_{pl}\sim M_{pl}$ (inflation ends before the
transition to the kination phase), and using (\ref{power1}) one obtains
$\bar{\lambda}\sim 10^{-14}$.

}

To end this Section, an important final remark is in order:
Taking into account that $x$ must be small -recall that $x$ is of the order of $\epsilon$- we can make, in formula (\ref{ns vs r}), the approximation $n_s\cong 1-3x$ and $r\cong 16x$ which leads, {}{for the potential (\ref{Pot})},  to the relation $n_s \cong 1-\frac{3}{16}r$. On the other hand, from the formula (\ref{A1}) for $n=2$, we will see that a quartic potential satisfy the same relation
$n_s \cong 1-\frac{3}{16}r$. Consequently, during inflation our family of potentials (\ref{Pot}) have the same behavior as a quartic potential, and  thus, the curve in the marginalized joint contour is the one that goes from  $62$ to $132$  e-folds in figure $1$.
Moreover, as we will see in the next Section, the reheating bounds due to the nucleosynthesis, i.e.,  the reheating temperature, must lie between $1$ MeV to a $10^9$ GeV (see \cite{gkr,hannestad}).
This leads, for our model, to a number of e-folds between $63$ and $73$, meaning that our
 models  (\ref{Pot}) are allowed, because they  enter in the $95.5 \%$ C.L. when the running is present.



\section{Reheating temperature in  quintessential inflation}
It is well-known that when the  potential has a minimum, the particles are created via the {}{inflaton}
 decay when it oscillates around the minimum \cite{kls}. On the contrary,
in {}{ quintessential inflation} the models do not have a minimum and it is usually assumed that gravitational particle production is due to an abrupt phase transition \cite{ford}, although there are other ways to  reheat the universe such as the so-called {\it instant preheating} \cite{fkl}.
{}{Moreover}, after the phase transition the
background energy density must decay faster {}{than} the energy density of the produced particles in order that, eventually, this last energy density  {}{dominates, which always happens if, after the phase transition, the universe enters in a kination regime.}


Here, I will study the reheating via gravitational creation of heavy massive particles {}{conformally coupled with gravity} during the phase transition.
Note
that, this is not the case studied in  early works, where
following the works of \cite{ford,Damour,Giovannini},
it was always assumed that the reheating was due to the production of very light particles \cite{Spokoiny,pv}.  In fact, is in the case of oscillating models, when
it is always assumed the
production of heavy massive particles due to the breakdown of the adiabatic regime during the oscillating regime.

{}{On the other hand, in our case the total Lagrangian of the system has the form ${\mathcal L}={\mathcal L}_G+{\mathcal L}_I+{\mathcal L}_{\bar\chi}$, where the gravitational
part is given
by ${\mathcal L}_G=\frac{a^3}{2}M_{pl}^2R$,  quintessential inflation is represented by ${\mathcal L}_I=\frac{a^3}{2}(\dot{\varphi}^2-2V(\varphi))$ and the corresponding part of the
massive quantum field
conformally coupled with gravity is ${\mathcal L}_{\bar\chi}=\frac{a^3}{2}(\dot{\bar{\chi}}^2-\frac{1}{a^2}|\nabla \bar{\chi}|^2-(m^2+\frac{R}{6})\bar{\chi}^2)$, where $R$ is the scalar
curvature and $m$ is the mass of the particles.

After the change of variable $\chi=a\bar{\chi}$ and working in the Fourier space, the Klein-Gordon equation corresponding to the quantum field acquieres the simple form of an harmonic
oscillator with a time dependent frequency
\begin{eqnarray}\label{wkb}
 \chi_k''+\omega_k^2(\tau)\chi_k=0,
\end{eqnarray}
where $'$ denotes the derivative with respect the conformal time $\tau$ and
$\omega_k(\tau)=\sqrt{k^2+m^2a^2(\tau)}$ is the frequency of the particle in the $k$-mode.

}

For the models presented in the previous Section,
the case $b=2$ is special because the second derivative of $H$ is nearly continuous, and in the other {}{cases ($b\not=2$)} is completely discontinuous.
This means,  that the number of created particles and consequently the reheating temperature will be greater in the case $b\not= 2$ than in
the exact case of a one-dimensional Higgs potential {}{combined with a cosmological constant}, because as we have already explained, particle production is due to the breakdown of the adiabaticity.
 Since the case $b=2$ has been studied in detail in \cite{he1}, {}{I} will concentrate in the case $b>2$.
Then,
 if during the adiabatic regimes, that is, when
 {}{\begin{eqnarray}\label{adiabatic}
 \omega_k'(\tau)\ll \omega_k^2(\tau)
 \Longrightarrow \sqrt{\lambda} M_{pl}\ll m \Longleftrightarrow H_e\ll m,\end{eqnarray}}
one uses the first order WBK solution {}{of (\ref{wkb})} to define approximately the vacuum modes \cite{Haro}
\begin{eqnarray}
\chi_{1,k}^{WKB}(\tau)\equiv
\sqrt{\frac{1}{2W_{1,k}(\tau)}}e^{-{i}\int^{\tau}W_{1,k}(\eta)d\eta},
\end{eqnarray}
where
\begin{eqnarray}
W_{1,k}=
\omega_k-\frac{1}{4}\frac{\omega''_{k}}{\omega^2_{k}}+\frac{3}{8}\frac{(\omega'_{k})^2}{\omega^3_{k}} .
\end{eqnarray}

{}{Then,
before the transition time, namely $\tau_E$ in conformal time,  the vacuum  state is depicted approximately by $\chi_{1,k}^{WKB}(\tau)$, but after the phase transition this mode becomes a mix of
positive and negative frequencies of the form
$\alpha_k \chi_{1,k}^{WKB}(\tau)+\beta_k (\chi_{1,k}^{WKB})^*(\tau)$, and 
the $\beta_k$-Bogoliubov coefficient, which is the key piece to calculate the number and energy density of the produced particles, could be obtained, as usual, matching both expressions at $\tau_E$, obtaining
\begin{eqnarray}\label{beta}
\beta_k=\frac{{\mathcal W}[\chi_{1,k}^{WKB}(\tau_E^-),\chi_{1,k}^{WKB}(\tau_E^+)]}
{{\mathcal W}[(\chi_{1,k}^{WKB})^*(\tau_E^+),\chi_{1,k}^{WKB}(\tau_E^+)]},
\end{eqnarray}
where $\tau_E^-$ (resp. $\tau_E^+$) means the limit on the left (resp. on the right) at the transition time, and
${\mathcal W}[f(\tau_E^-),g(\tau_E^+)]\equiv f(\tau_E^+)g'(\tau_E^-)-f'(\tau_E^+)g(\tau_E^-)$ is the Wronskian of the functions $f$ and $g$ at the transition time.

}

Then,
the square modulus of the {}{$\beta_k$-Bogoliubov} coefficient will be given by (see {}{\cite{he2}} for a detailed discussion about the calculation of the number of
created particles){}{
\begin{eqnarray}
 |\beta_k|^2\cong \frac{m^4a_E^{10}\left(\ddot{H}_E^+-\ddot{H}_E^-\right)^2}{256(k^2+m^2a^2_E)^5},
\end{eqnarray}
where $\ddot{H}_E^-$ (resp. $\ddot{H}_E^+$)},
is the value of the second derivative of the Hubble parameter before (after) the phase transition, and $a_E$
denotes the value of the scale factor at the phase transition time.
The number density of produced particles and their energy density is {}{\cite{birrell}}
\begin{eqnarray}
n_{\chi}\equiv \frac{1}{2\pi^2 a^3}\int_0^{\infty}k^2|\beta_k|^2dk,\qquad 
\rho_{\chi}\equiv \frac{1}{2\pi^2 a^4}\int_0^{\infty}\omega_k k^2|\beta_k|^2dk,
\end{eqnarray}
{}{therefore, since for the model (\ref{Pot}) we have the analytic  background (\ref{background}), we can calculate explicitly the second
derivative of the Hubble parameter  $\ddot{H}_E^-\cong 9bH_e^3$ and  $\ddot{H}_E^+\cong 18H_e^3$ ,
leading for the potential  (\ref{Pot})}, the following number and energy density
\begin{eqnarray}
 n_{\chi}\sim 10^{-3}(b-2)^2\frac{H_e^6}{m^3}\left(\frac{a_E}{a} \right)^3, \quad \rho_{\chi}\sim mn_{\chi}.
\end{eqnarray}

{}{ Unfortunately, for the potential (\ref{Pot1})  there is not any analytic background, then to calculate the second derivative
of the Hubble parameter, I will use the equation
\begin{eqnarray}
\dot{H}=-\frac{\dot{\varphi}^2}{2M_{pl}^2}\Longleftrightarrow \ddot{H}=-\frac{\dot{\varphi}\ddot{\varphi}}{M_{pl}^2}.
\end{eqnarray}

Note that, from the conservation equation, at the transition time, one has
\begin{eqnarray}
|\ddot{\varphi}(t_E^-)-\ddot{\varphi}(t_E^+)|=|V_{\varphi}(M_{pl}^-)|=4\bar{\lambda}M_{pl}^3.
\end{eqnarray}

To calculate $\dot{\varphi}(t_E)$, I use that at the transition time all the energy is kinetic, which means that
$|\dot{\varphi}(t_E)|=\sqrt{6}H_E M_{pl}$, where $H_E$ is the value of the Hubble parameter at the transition time. The problem is that, since we do not have a background, one cannot calculate that value. Therefore, some assumptions must be made, for example one can argue that there is no substantial drop of energy from the last stages of inflation to the transition, or just to take, as for the similar potential (\ref{Pot}) where there exists  an analytic background , $H_E\sim H_e\sim 10^{-7}M_{pl}$.

Considering the second case we get
\begin{eqnarray}
|\ddot{H}_E^--\ddot{H}_E^+|\sim 10^{-7}\bar{\lambda}M_{pl}^3\sim H_e^3,
\end{eqnarray}
and we will obtain a number density of produced particles  of the same order as for the potential (\ref{Pot}).
}

These {}{non relativistic} particles  are far from being in thermal equilibrium {}{and, at the beginning, their energy density scales
as $a^{-3}$, eventually}  they will decay into lighter
particles, which will interact through multiple scattering. Then, a re-distribution of energies among the different
particles occurs
and also, an increase in the number of particles will occur \cite{37}. {}{At the end of these process, the universe becomes filled by a
relativistic plasma in thermal equilibrium whose energy density decays as $a^{-4}$. Then, in order to obtain the reheating temperature, first of all, one needs to calculate the moment when thermalization occurs, because there is a first  period where the energy density of the produced particles scales as matter, and another one, after thermalization,  where it scales as radiation. }

{}{To do that}, I will use the thermalization process depicted in \cite{38}, where the cross section for $2\rightarrow 3$ scattering with  gauge bosons exchange whose typical energy is $\rho_{\chi}^{\frac{1}{4}}(t_E) $ is given by
$\sigma={\alpha^3}\rho_{\chi}^{-\frac{1}{2}}(t_E)$, with $\alpha^2\sim 10^{-3}$. Then, the thermalization rate is
\begin{eqnarray}
 \Gamma=\sigma n_{\chi}(t_E)
 \sim 10^{-3/2}(b-2)\alpha^3\left(\frac{H_e}{m}\right)^2H_e.
\end{eqnarray}
  Equilibrium is reached when {}{$\Gamma\sim H(t_{eq})\cong H_e\left(\frac{a_E}{a_{eq}}\right)^3$} (recall that after the phase transition the background evolves like it was
  driven by stiff matter),  which leads to the relation {}{$\frac{a_E}{a_{eq}}\sim 10^{-1/2}(b-2)^{1/3}\alpha \left(\frac{H_e}{m}\right)^{2/3}$}. Then, at the equilibrium one has
\begin{eqnarray}
 \rho_{\chi}(t_{eq})\sim 10^{-9/2}(b-2)^3\alpha^3\left(\frac{H_e}{m}\right)^{4}H_e^4,\nonumber\\
 \rho(t_{eq})\sim 3\times 10^{-3}(b-2)^2\alpha^6\left(\frac{H_e}{m}\right)^{4}H_e^2 M_{pl}^2.
\end{eqnarray}

After this thermalization, the relativistic plasma evolves as {}{$\rho_{\chi}(t)=\rho_{\chi}(t_{eq})\left(\frac{a_{eq}}{a} \right)^4$, and the background evolves as
$\rho(t)=\rho(t_{eq})\left(\frac{a_{eq}}{a} \right)^6$}, because we are in the deflationary regime. The reheating is obtained when both energy densities are 
of the same order, {{}and that} will happen when {}{$\frac{a_{eq}}{a_R}\sim \sqrt{\frac{\rho_{\chi}(t_{eq})}{\rho(t_{eq})}}$}, and thus,
obtain the reheating temperature of the order
\begin{eqnarray}
 T_R\sim \rho_{\chi}^{\frac{1}{4}}(t_{eq})\sqrt{\frac{\rho_{\chi}(t_{eq})}{\rho(t_{eq})}} \sim 
 10^{-1}\left(\frac{H_e}{M_{pl}} \right)^2\left(\frac{H_e}{m}\right)M_{pl}.
\end{eqnarray}

{}{
Since, as I have showed in (\ref{lambda}), $H_e\sim 10^{-7} M_{pl}$ and I am considering heavy massive particles satisfying the condition
(\ref{adiabatic}), we have to choose particles with
 masses of the order $ 10^{13}$ GeV or greater. The greater reheating temperature,
$T_R\sim 10$ GeV, is achieved when $ m\sim 10^{13}$ GeV.

To calculate, for our models, the number of e-folds I use the formula (\ref{number}) with $n=\infty$.
Taking  $n_s=0.96$, $g_R=90$ and $T_R\sim 10$ GeV one obtains $N(k_*)\cong 68$.
Moreover, integrating (\ref{background}), after the phase transition one has $H(t)=\frac{H_E}{1+3H_E(t-t_E)}$, and then, at the reheating
time $H_R=\frac{H_E}{1+3H_E(t_R-t_E)}\sim \frac{1}{t_R-T_E}$ when $H_E(t_R-t_E)\gg 1$. Since $H_R\sim \frac{T_R^2}{M_{pl}}\sim 10^{-34}
M_{pl}$, and using that $t_{pl}\sim 10^{-44}$ s, one obtains that the universe reheats around $10^{-10}$ s after the phase transition.
}


\vspace{1cm}
To understand the fate of the universe, we have to realize that when it
 becomes reheated it is filled by a thermalized relativistic plasma  whose energy density is greater than that of the background, and thus, 
this relativistic plasma  drives the evolution of the universe in the same way as in the $\Lambda$CDM model. Then, since the energy density of this plasma decays as
$a^{-4}$ during the radiation domination and as $a^{-3}$ when matter starts to dominate, eventually the energy density of the field will dominate because the potential
energy $V(\varphi)=M_{pl}^2\Lambda$ is constant. In fact, if we choose $\Lambda\sim H_0^2$ the kinetic energy of the field will becomes sub-dominant which respect the potential one. Effectively,
since after the phase transition the potential is constant one will have
\begin{eqnarray}
 \ddot{\varphi}+3H\dot{\varphi}=0\Longleftrightarrow \dot{\varphi}(t)=\dot{\varphi}(t_R)e^{-3\int_{t_R}^tH(s)ds},
\end{eqnarray}
where $t_R$ is the reheating time.

On the other hand, during the radiation and the matter dominated phases,  one will have
\begin{eqnarray}
 H(t)=\frac{H_R}{1+2(t-t_R)H_R}, \quad \mbox{and} \quad H(t)=\frac{2H_M}{2+3(t-t_M)H_M},
\end{eqnarray}
where the subindices $R$, $M$ respectively denote the Hubble rate when radiation and matter domination will start to dominate. Then, at $t=t_0$, one will get
\begin{eqnarray}
 \dot{\varphi}(t_{0})=\frac{\dot{\varphi}(t_R)}{(1+2(t_M-t_R)H_R)^{\frac{3}{2}}(2+3(t_{0}-t_M)H_M)^2}\Longrightarrow 
 \dot{\varphi}^2(t_{0})\sim \dot{\varphi}^2(t_R)\frac{H_MH_0^2}{H_R^3},
\end{eqnarray}
}
and consequently,  since at the beginning of the radiation era all the energy of the field is kinetic one will have $ \dot{\varphi}^2(t_R)\sim M_{pl}^2 H^2_R$, meaning that
 nowadays the ratio between the kinetic and potential energy density, namely  $\mathcal{R}$,  satisfies
\begin{eqnarray}
\mathcal{R}\cong \frac{\dot{\varphi}^2(t_0)/2}{\Lambda M_{pl}^2}\sim \frac{H_M}{H_R}\sim \left(\frac{T_M}{T_R}\right)^2\ll 10^{-20},
\end{eqnarray}
where I've used that for our models the reheating temperature is around $10$ GeV, and the temperature at the beginning of the matter domination, as we have seen in Section II, is given by $T_M=\frac{a_0}{a_M}T_0$, with  $T_0\cong 2.4 \times 10^{-13}$ GeV and $\frac{a_0}{a_M}=3360$ \cite{rg}.

Summing up, if one chooses $\Lambda\sim H_0^2$ nowadays 
the kinetic part of the energy density of the field is sub-dominant.
Moreover, the energy density of the matter is of the order $M_{pl}^2H_0^2\sim V(\varphi)$, which means that 
 it is the cosmological constant which drives the current evolution of the universe. And thus, the Friedmann equation becomes
 $H^2\cong\frac{\Lambda}{3M_{pl}^2}$, showing that the universe accelerates with and effective Equation of State parameter nearly $-1$.

\vspace{1cm}


{}{Three final remarks are in order:}
\begin{enumerate}
\item
 For {}{$b>2$, the  first derivative of the potential is discontinuous,} this implies that $\ddot{\varphi}$ and consequently {}{$\ddot{H}$} are discontinuous at the transition time. However, for $b=2$,
is the third derivative of the Hubble parameter which is discontinuous at the transition time, leading to an smaller temperature in the regime of the MeV {}{\cite{he1}}. 
On can argue that in all cases, i.e. for $b\geq 2$, one has a past Type IV singularity, called "Big Brake" in \cite{Lazkoz} and "Generalized Sudden" in \cite{BHO}, because
the energy density $\rho=3M_{pl}^2H^2$ and the pressure $P=-2M_{pl}^2\dot{H}+3M_{pl}^2H^2$ are continuous all the time, but higher derivatives of the Hubble parameter are divergent at the transition time. In fact, for $b>2$ the second derivative of $H$ is discontinuous what means -in a mathematical distributional sense- that the third derivative of $H$ diverges at the transition time. In the same way, one can see that for $b=2$ is the fourth derivative which diverges at the transition time.
However here, it is important to realize that this singularity appears because in the models I have assumed an instantaneous phase transition in order to have a simple way to calculate the Bogoliubov coefficients (see formula (\ref{beta})), but if one assumes a phase transition that  last a short period of time,  then that undesirable behavior is removed because
the derivatives of the Hubble parameter do not diverge. The difficulty when the phase transition is not instantaneous  is that it is very complicated to calculate the Bogoliubov coefficients,  this is the reason why  I have considered an instantaneous one. 

\item Even worse happens with the model proposed by Peebles and Vilenkin in \cite{pv}, 
where the Hubble parameter is continuous up to the fourth derivative, what leads to an abnormally small temperature in the eV regime.
{}{And the same happens with potentials that have a more abrupt phase transition
\begin{eqnarray}
V(\varphi)=\left\{\begin{array}{ccc}
\bar{\lambda}(\varphi^4+M^4)&\mbox{when}& \varphi<0\\
\frac{\bar{\lambda}M^4}{(\varphi/M)^{n}+1}&\mbox{when}& \varphi>0,\end{array}
\right.
\end{eqnarray} with $n=1,2,3$. A similar calculation as the one performed
above shows that the gravitational production of heavy  massive particles leads to an abnormally small reheating temperature in the eV regime.

\item
In \cite{39} the authors consider exponential potentials of the form  
$V(\varphi)=V_0e^{-\lambda \varphi^n/M_{pl}^n}$, showing that, without allowing the running, for several values of the parameters $\lambda$ and $n$
the model lies inside  the $68\%$ C.L. of the marginalized  joint coincidence contour in the plane $(n_s,r)$, and clearly improve the results obtained in this work. However, the authors 
argue that the reheating temperature is greater than $10^{14}$ GeV, which not seems to be in agreement with nucleosynthesis bounds.}
\end{enumerate}

\section{Accurate calculation of the number of e-folds for some quintessential potentials}
For the {}{quintessential inflation} models presented in the previous section we have an explicit expression of 
the 
\begin{eqnarray}
\frac{k_*}{a_0H_0}
=e^{-N(k_*)}\frac{H(k)}{H_0}\frac{a_{end}}{a_E}\frac{a_E}{a_R}\frac{a_R}{a_M}\frac{a_M}{a_0}=
e^{-N(k_*)}\frac{H(k_*)}{H_0}\frac{a_{end}}{a_E}
\frac{\rho_R^{-1/12}\rho_M^{1/4}}{\rho^{1/6}_E}\frac{a_M}{a_0},
\end{eqnarray}
where once again, $t_E$ is the phase transition time,
and we have used the relation between the scale factor and the energy density in the corresponding different
phases.

Then, if we choose the pivot scale as $k_*=0.005$ $\mbox{Mpc}^{-1}$,
 one will have
\begin{eqnarray}
N(k_*)=-5.52+\ln\left(\frac{H(k_*)}{H_0} \right)+\ln\left(\frac{a_{end}}{a_E} \right)
+\frac{1}{4}\ln\left(\frac{\rho_M}{\rho_R} \right) +\frac{1}{6}\ln\left(\frac{\rho_R}{\rho_E} \right)
+\ln\left(\frac{a_M}{a_0} \right).
\end{eqnarray}

To calculate $\ln\left(\frac{a_{end}}{a_E} \right) $, we will use (\ref{end}), to obtain
{}{\begin{eqnarray}
\ln\left(\frac{a_{end}}{a_E} \right)\cong\int_{H_e}^{H_{end}}\frac{H}{\dot{H}}dH=
-\frac{b-1}{3b^2}\ln\left[\frac{3b}{2}\left(1+\sqrt{1-\frac{4(b-1)}{3b^2}} \right)-\frac{b-1}{b}   \right]\nonumber\\
-\frac{1}{3b}\left[\frac{3b}{2}\left(1+\sqrt{1-\frac{4(b-1)}{3b^2}} \right)-1 \right]
\cong -1,
\end{eqnarray}
for $b \sim {\mathcal O(1)}$.}

Using  $\frac{a_0}{a_M}=3360$ \cite{rg}, one obtains
$\ln\left(\frac{a_M}{a_0} \right)\cong -8.12$. Further, from (\ref{power1})
we also have
\begin{eqnarray}
\ln\left(\frac{H(k_*)}{H_0}\right)\cong 130.83 +\frac{1}{2}\ln\left(\frac{1-n_s}{3} \right)\cong 128.67.
\end{eqnarray}
Now, since the current temperature of the cosmic background is $T_0= 2.73$ K and the
conservation of entropy implies $T_M=\frac{a_0}{a_M}T_0\sim 9\times 10^3$ K $\sim 9\times 10^{-10}$ GeV,  using that
$\frac{\rho_M}{\rho_R} \cong \frac{2g_MT_M^4}{g_RT_R^4}  $, we obtain that
\begin{eqnarray}
\frac{1}{4}\ln\left(\frac{\rho_M}{\rho_R} \right)\cong -20.35-\ln\left(\frac{g_R^{1/4}T_R}{\mbox{GeV}}  \right).
\end{eqnarray}
Moreover, for our model it turns out that $\rho_E^{\frac{1}{4}}\cong 1.63 {\lambda}M_{pl}\sim 4\times 10^{14}$ GeV. Then,
\begin{eqnarray}
\frac{1}{6}\ln\left(\frac{\rho_R}{\rho_E} \right)=\frac{2}{3}\left(-33.89+
\ln\left(\frac{g_R^{1/4}T_R}{\mbox{GeV}}  \right)
\right).
\end{eqnarray}

Finally, collecting  all the results above, it follows that
we obtain
{}{\begin{eqnarray}
N(k_*)\cong 71.07-\frac{1}{3}\ln\left(\frac{g_R^{1/4}T_R}{\mbox{GeV}}  \right),
\end{eqnarray}}
which means that if the reheating temperature --with the purpose to ensure the success of nucleosynthesis--  needs to belong in the range between $10^9$ GeV and $1$ MeV, then
 the number of e-folds must lie between $63$ and $73$. In particular, when the reheating temperature is of the order of $10$ GeV --the scale we obtain
 if reheating is due to the creation of heavy particles with masses of about $10^{13}$ GeV during the phase transition--
 the number of e-folds of the universe expansion in our model
 is approximately
 {}{$70$}, which perfectly enters in the
$95.5 \%$ C.L. when the running is allowed, because the model gives the same results as the quartic potential (curve between $62$ and 
$132$ e-folds in figure $1$).

\vspace{0.5cm}

\section{Discussion}
\label{discuss}
Some  simple quintessential  inflation models, which are the combination of the Higgs potential with a cosmological constant,
are presented. For them,
the reheating temperature {due to the creation of heavy massive particles during the phase transition, is approximately  $10$ GeV,  the number of e-folds around $68$}
and the theoretical values of 
{}{$n_s$ and $r$}
enter in the $95.5 \%$ C.L. of the two dimensional marginalized joint coincidence  contour for $(n_s,r)$ when the running is allowed, showing the models' viability.

\vspace{0.5cm}

\acknowledgments
I would like to thank Professor Jaume Amor\'os for reading carefully the manuscript and helping in the numeric calculations performed in  figure $1$, and to the anonymous referee for the comments that have been very useful to improve this work.
This investigation has been supported in part by MINECO (Spain), project MTM2014-52402-C3-1-P.

\end{document}